\documentclass[conference]{IEEEtran}
\usepackage{algorithm,algorithmic}
\usepackage{amssymb}
\usepackage{amsmath}
\usepackage{dsfont}
\usepackage{cite}
\usepackage{url}
\usepackage{xcolor}
\usepackage{cite,graphicx,amsmath,amssymb}
\usepackage{fancyhdr}
\usepackage{mdwmath}
\usepackage{mdwtab}
\usepackage{amsthm}
\usepackage{multirow}
\usepackage{flafter}
\usepackage{cite,graphicx,amsmath,amssymb}
\usepackage[lofdepth,lotdepth]{subfig}
\usepackage{fancyhdr}
\usepackage{mdwmath}
\usepackage{mdwtab}
\usepackage{balance}
\usepackage{xcolor}
\usepackage{bm}
\usepackage{url}
\usepackage{float}
\usepackage{caption}
\usepackage{amsthm}
\usepackage{algorithm}
\usepackage{algorithmic}
\usepackage{multirow}
\usepackage{flafter}
\usepackage{setspace}

\renewcommand{\algorithmicrequire}{\textbf{Input:}} %
\renewcommand{\algorithmicensure}{\textbf{Return:}}
\usepackage{amsmath,amsthm,amssymb}

\usepackage[
top    = 1.71cm,
bottom = 1.05in,
left   = 0.63 in,
right  = 0.63 in]{geometry}

\newtheorem{remark}{Remark}
\newtheorem{theorem}{Theorem}

\newtheorem{lemma}{Lemma}

\newtheorem{corollary}{Corollary}

\makeatletter
\def\ScaleIfNeeded{%
\ifdim\Gin@nat@width>\linewidth \linewidth \else \Gin@nat@width
\fi } \makeatother

\IEEEoverridecommandlockouts
\begin{document}

\title{Meta-learning for RIS-assisted NOMA Networks}
\author{
\IEEEauthorblockN{ 
Yixuan~Zou\IEEEauthorrefmark{1},Yuanwei~Liu\IEEEauthorrefmark{1},Kaifeng~Han\IEEEauthorrefmark{2},Xiao~Liu\IEEEauthorrefmark{1},Kok~Keong~Chai\IEEEauthorrefmark{1}\\}
\IEEEauthorblockA{
\IEEEauthorrefmark{1} Queen Mary University of London, London, UK\\
\IEEEauthorrefmark{2} China Academy of Information and Communications Technology, Beijing, China\\
\{yixuan.zou, yuanwei.liu, x.liu, michael.chai\}@qmul.ac.uk\\
hankaifeng@caict.ac.cn
}
}
\maketitle

\begin{abstract}
A novel reconfigurable intelligent surfaces (RISs)-based transmission framework is proposed for downlink non-orthogonal multiple access (NOMA) networks. We propose a quality-of-service (QoS)-based clustering scheme to improve the resource efficiency and formulate a sum rate maximization problem by jointly optimizing the phase shift of the RIS and the power allocation at the base station (BS). A model-agnostic meta-learning (MAML)-based learning algorithm is proposed to solve the joint optimization problem with a fast convergence rate and low model complexity. Extensive simulation results demonstrate that the proposed QoS-based NOMA network achieves significantly higher transmission throughput compared to the conventional orthogonal multiple access (OMA) network. It can also be observed that substantial throughput gain can be achieved by integrating RISs in NOMA and OMA networks. Moreover, simulation results of the proposed QoS-based clustering method demonstrate observable throughput gain against the conventional channel condition-based schemes.
\end{abstract}


\section{Introduction}
Multiple-input-multiple-output (MIMO) has been recognized as a promising technology to enhance the capacity and the spectral efficiency of fifth-generation (5G) wireless networks~\cite{MIMO}. However, deploying and configuring a large number of antennas can lead to severe hardware impairment, heavy computational cost, and substantial power consumption. To overcome these limitations, reconfigurable intelligent surfaces (RISs) have emerged as a cost-effective solution~\cite{RIS_review_Liu21}. An RIS composes of a large number of low-cost reflecting elements that can proactively reconfigure the propagation of incident signals. By intelligently adjusting the phase shift of each RIS element, the communication channels can be effectively manipulated to enhance the spectral efficiency and the network coverage~\cite{Xiao_UAV-RIS, Xiao_NOMA_RIS}. Benefiting from the low-cost meta-materials, RISs can be seamlessly integrated with emerging technologies, such as MIMO, to further improve transmission throughput in a cost-effective manner.

As a promising technology for supporting massive connectivity, non-orthogonal multiple access (NOMA)~\cite{NOMA_yuanwei} can be integrated into RIS-aided networks to further enhance the spectral efficiency. Moreover, the integration of NOMA and RIS can effectively improve the design flexibility of NOMA schemes~\cite{RIS_NOMA_interplay_Liu20}. However, the implementation of NOMA in RIS-aided networks increases the resource allocation difficulty. In particular, RISs can alter the channel quality of individual devices, which directly influences the decoding orders and the clustering results of NOMA systems.

Extensive research contributions have been devoted to investigating the integration of RIS and NOMA techniques~\cite{SoA_RIS_NOMA_Hou20, SoA_RIS_NOMA_Mu20, SoA_RIS_NOMA_Yang20}. In particular, the authors of~\cite{SoA_RIS_NOMA_Hou20} derived extensive analytical results, including ergodic rates, energy efficiency and spectral efficiency, for RIS-assisted NOMA networks. The sum rate maximization problem of RIS-aided NOMA systems was investigated in~\cite{SoA_RIS_NOMA_Mu20}, where the passive beamforming at the RIS was jointly optimized with the active beamforming at the base station (BS) under both the ideal and non-ideal RIS elements. To maximize sum rate while ensuring user fairness, the authors of~\cite{SoA_RIS_NOMA_Yang20} formulated a max-min problem for RIS-enhanced NOMA networks by jointly optimizing the power allocation and the RIS phase shift. The aforementioned research contributions revealed the potentials of RIS when integrated into NOMA networks and established a foundation for solving various challenges in RIS-assisted networks. However, these contributions mainly investigated the implementations of conventional convex optimization techniques, which often suffer from high computational complexity and poor scalability. Moreover, the objective functions are often non-convex, which can not be directly tackled by convex optimization methods. Hence, alternative cost-effective non-convex optimization schemes are necessary to fulfil the requirements of massive connectivity in next-generation wireless networks.

In recent years, artificial intelligence (AI) has emerged as a tremendous technology to address the problems of exploding data volume, non-convex optimization, and computational complexity~\cite{ML_survey1}. In particular, deep learning (DL) techniques utilize the extensive offline training phase to reduce the algorithm complexity in the application and have received overwhelming research interests in the optimization of RIS-assisted wireless communication systems~\cite{RIS_BF_opt_transferDL_Ge,  RIS_indoor_DL_Huang, RIS_DL_PS_Sheen}. In~\cite{RIS_BF_opt_transferDL_Ge}, deep transfer learning was employed to solve the beamforming optimization problem in multiple-input-single-output (MISO) networks, based on a small amount of training data. The problem was further extended into the discrete phase shift cases to accommodate hardware limitations. The authors of~\cite{RIS_indoor_DL_Huang} and~\cite{RIS_DL_PS_Sheen} utilized neural networks to learn the interactions between the receiver locations and the optimal RIS phase shift to achieve maximal communication throughput. These aforementioned contributions demonstrated the outstanding performance of DL-based techniques when solving high-dimensional and non-convex optimization problems in RIS-enhanced wireless networks. 

However, existing DL-based RIS optimization methods~\cite{RIS_BF_opt_transferDL_Ge, RIS_indoor_DL_Huang, RIS_DL_PS_Sheen} are all based on orthogonal multiple access (OMA) systems, where the RIS phase shift is the only optimization variable. To the best of our knowledge, there does not exist a DL-based solution for RIS-aided NOMA networks, which motivates this study. 
In this paper, we investigate the sum rate optimization problem in RIS-aided downlink MISO-NOMA networks, where both the RIS phase shift and the BS power allocation are optimized to maximize the total transmission sum rate. We adopt the zero-forcing (ZF) precoding method and the successive interference cancellation (SIC) decoding method to eliminate the effect of multi-user interference on the strong users. However, this approach causes the weak users to suffer from both inter-cluster and intra-cluster interference, leading to poor achievable rates. To improve the resource efficiency, we propose a quality-of-service (QoS)-based NOMA clustering method, which aims to maximize the QoS deviation within each cluster. In terms of the DL model, we design a neural network to output optimized power allocation given the RIS phase shift, resulting in a low-complexity model. Meanwhile, the phase shift is optimized through a gradient descent algorithm, given the trained network. We further employ meta-learning in the training process to improve the convergence rate of the phase shift optimization. The main contributions are as follows:
\begin{enumerate}
    \item We propose a RIS-enhanced NOMA downlink framework and formulate the sum rate maximization problem by jointly optimizing the phase shift of the RIS and the power allocation of the BS. To improve the resource efficiency, we propose a QoS-based NOMA clustering scheme, which maximizes the QoS deviation within clusters.
    \item We propose a MAML-based DL algorithm to solve the joint optimization problem. The algorithm can output optimized solutions in as few as five iterations and the model has lower network complexity compared to the conventional design.
    \item Simulation results indicate that the implementation of RIS can induce approximately 5\% to 25\% throughput gain as the number of RIS elements increases from 8 to 64, in both NOMA and OMA networks. Results also show that the proposed QoS-based clustering method achieves higher throughput than the conventional channel condition-based approach.
\end{enumerate}

\section{System Model and Problem Formulation}\label{sec:system model}
\subsection{System Model}
\begin{figure}[t]
    \includegraphics[width=0.47\textwidth]{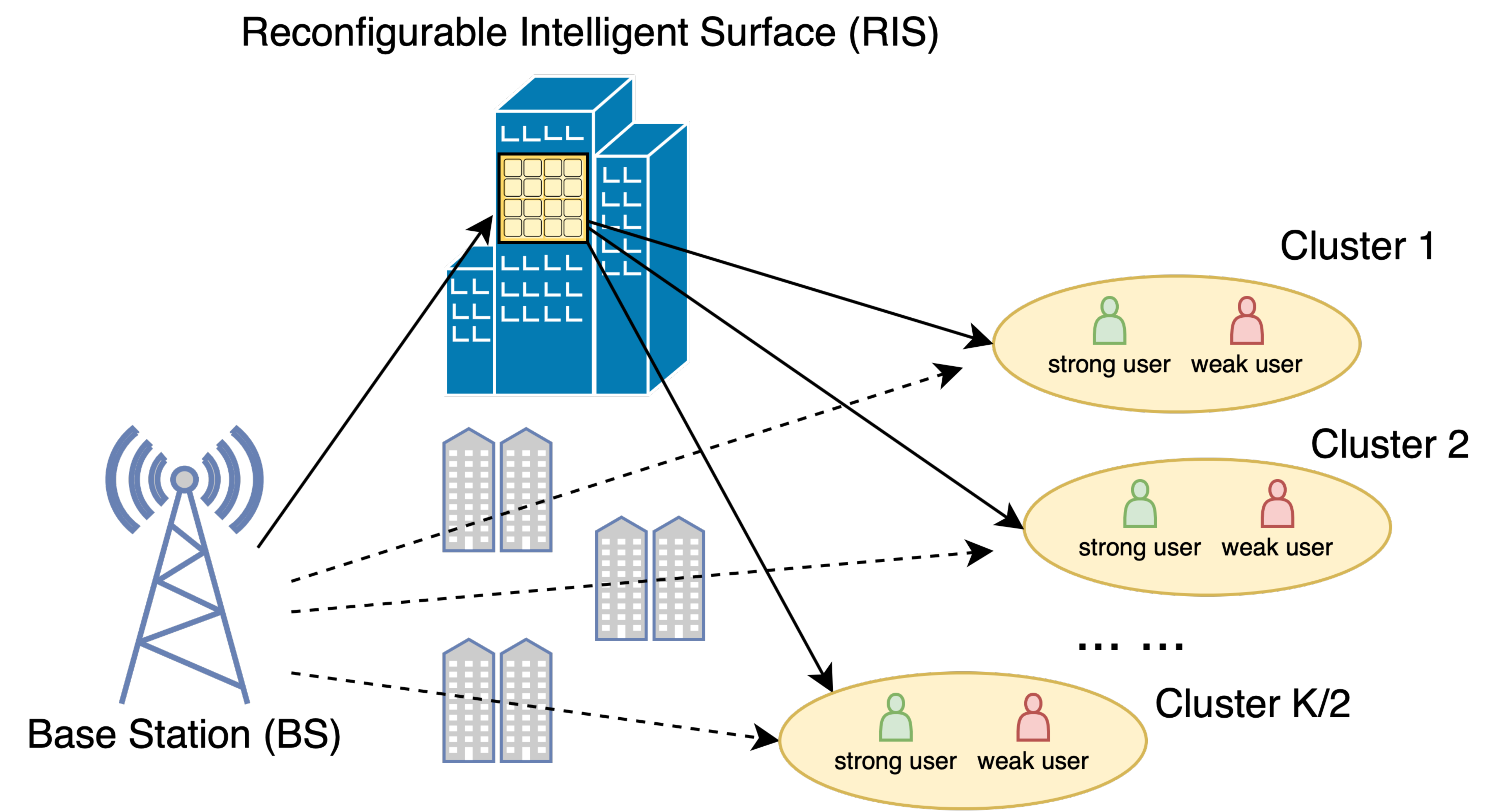}
    \caption{Illustration of the RIS-assisted downlink MISO-NOMA network.}
    \label{fig: system model}
\end{figure}
As illustrated in Fig.~\ref{fig: system model}, we consider a downlink MISO system with one BS and $K$ mobile users (MUs). The BS is equipped with $M$ antenna elements and each MU is equipped with one single antenna. The communication between the BS and the MUs is assisted by a RIS of $N$ reflecting elements, whose phase shift and amplitude absorption can be adjusted by a controller.

The channels between the BS and the RIS are modelled as Rician fading channels. The BS-MU channels and the RIS-MU channels are modelled as Rayleigh fading channels. The path loss of a particular MU $k$ is modelled as $\text{PL}_k = d_k^{-\alpha}$ where $d_k$ is the distance, calculated in meters, between the MU and the BS, and $\alpha$ denotes the path loss exponent.


\subsection{NOMA Signal Model}
In this subsection, we formulate the NOMA-based signal model and introduce the proposed QoS-based clustering method.

\subsubsection{Signal model}
The signal received at each MU is a composition of the signals derived from the direct link between BS and MU, and the signals derived from the reflecting link. In particular, for MU $i$ in the $l$-th cluster, we denote the RIS-MU link and BS-MU link by  $\mathbf{h}_{R,l,i}^{H} \in {\mathbb{C}^{1 \times N}}$, and $\mathbf{h}_{B,l,i}^{H} \in {\mathbb{C}^{1 \times M}}$, respectively. We further denote the BS-RIS link by ${\mathbf{H}}_{BR} \in {\mathbb{C}^{N \times M}}$. The phase shift of the RIS is denoted by $\boldsymbol{\theta} = [\theta_1, \cdots, \theta_n, \cdots, \theta_N]$ where $\theta_n \in [0, 2\pi)$. The diagonal phase-shifting matrix is expressed as $\mathbf{\Theta} = \text{diag}(\beta_1 e^{j\theta_1},\cdots,\beta_n e^{j\theta_n},\cdots,\beta_N e^{j\theta_N})$, where $\beta_n \in [0,1]$ is the amplitude reflection coefficient. For simplicity, we assume that all amplitude coefficients are ones, i.e., $\beta_n = 1, \forall n$.

We assume that each cluster is formed by two users and we denote the strong MU as MU $s$ and the weak MU as MU $w$. The signals transmitted to the strong and the weak MU in the $l$-th cluster are denoted by ${s_{l,s}}$ and ${s_{l,w}}$, respectively and we denote the transmit power allocated to the strong and the weak MUs in the $l$-th cluster by $p_{l,s}$ and $p_{l,w}$, respectively. Hence, the transmit signal of the $l$-th cluster is formulated as ${x_l} = \sqrt {{p_{l,s}}} {s_{l,s}} + \sqrt {{p_{l,w}}} {s_{l,w}}$. The corresponding signal received by MU $i$ in the $l$-th cluster can be expressed as
\begin{align}\label{transmitsignalNOMA}
{{y_{l,i}} = \left( {{\mathbf{h}}_{B,l,i}^H + {\mathbf{h}}_{R,l,i}^H{{\mathbf{\Theta }}}{{\mathbf{H}}_{BR}}} \right)\sum\limits_{l = 1}^{K/2} {{{\mathbf{w}}_l}{x_l}}  + {n_{l,i}} },
\end{align}
where ${{{\mathbf{w}}_l}}$ denotes the beamforming vector of the $l$-th cluster and $n_{l,i}$ denotes the additive white Gaussian noise (AWGN), modelled as $n_{l,i} \sim \mathcal{C}\mathcal{N}\left( {0,\sigma ^2} \right)$. 

To decode the received symbols from the multiplexed signal $x_l$, each strong MU employs the SIC technique which eliminates the intra-cluster interference. The weak MU, however, decodes the signal directly without SIC. Additionally, we eliminate the inter-cluster interference for the strong MUs through the ZF beamforming technique. 
The normalized ZF precoding vector is given by
\begin{align}\label{ZFNOMA}
{\mathbf{w}_{l}} = \frac{{\mathbf{h}_{l,s}}{\left( {{{\mathbf{h}^H_{l,s}}}{\mathbf{h}_{l,s}}} \right)^{ - 1}}}{\rho_{l,s}},
\end{align}
where $\mathbf{h}_{l,s}^H = \mathbf{h}^H_{B,l,s} + \mathbf{h}^H_{R,l,s}\mathbf{\Theta}\mathbf{H}_{BR}$ denotes the combined channel and  $\rho_{l,s}$ denotes the normalizing constant formulated as $\rho_{l,s} = |{{\mathbf{h}_{l,s}}{( {{{\mathbf{h}_{l,s}}^H}{\mathbf{h}_{l,s}}} )^{ - 1}}} |^2$. The corresponding ZF precoding constraints are expressed as follows
\begin{align}\label{ZF2}
\left\{ \begin{array}{*{20}{c}}
  \mathbf{h}_{j,s}^H \mathbf{w}_l &= 0, & {\kern 1pt} \forall j \ne l,{\kern 1pt} {\kern 1pt} {\kern 1pt} j = 1, \cdots, {K/2} , \\
  \mathbf{h}_{j,s}^H \mathbf{w}_l&= \frac{1}{\rho_{l,s}}, & j = l.
\end{array} \right.
\end{align}

Hence, without the interference, the signal received at the strong MU in the $l$-th cluster can be simplified into
\begin{align}\label{receiveNOMA}
{{y_{l,s}}{\kern 1pt}  = \mathbf{h}^H_{l,s}{{\mathbf{w}}_l}\sqrt {{p_{l,s}}} {s_{l,s}} + {n_{l,s}}}.
\end{align}

Based on \eqref{ZF2}, we can calculate the received SINR of the strong MU in the $l$-th cluster as
\begin{align}\label{eq: SINRNOMA}
{{\gamma _{l,s}} = \frac{{{{\left| {\mathbf{h}^H_{l,s}{{\mathbf{w}}_l}\sqrt {{p_{l,s}}} {s_{l,s}}} \right|}^2}}}{{\sigma ^2}} = \frac{p_{l,s}}{{\rho_{l,s}\sigma ^2}} }.
\end{align}

Since both inter-cluster interference and intra-cluster interference exists in the weak MU's received signals, the received SINR of the weak MU in the $l$-th cluster is derived as
\begin{align}\label{eq: weak user SINR}
{{\gamma _{l,w}} = \frac{{{{\left| {{{\mathbf{h}}_{l,w}}{{\mathbf{w}}_l}} \right|}^2}{p_{l,w}}}}{{{{\left| {{{\mathbf{h}}_{l,w}}{{\mathbf{w}}_l}} \right|}^2}{p_{l,s}} + {{\left| {{{\mathbf{h}}_{l,w}}\sum\limits_{j = 1,j \ne l}^{K/2} {{{\mathbf{w}}_j}{x_j}} } \right|}^2} + \sigma ^2}} }.
\end{align}

\subsubsection{QoS-based clustering scheme}
When both the ZF precoding and the SIC decoding techniques are employed in NOMA, the weak MUs suffer from both the inter-cluster and the intra-cluster interference, resulting in low SINR and low achievable rate compared to the strong MUs, who are served in an interference-free manner. Conventional clustering methods allocate MUs by exploiting the difference between their channel conditions. However, when MUs have different QoS requirements, we may notice a weak MU acquiring a high QoS, which is challenging to achieve given the multiuser interference. Moreover, due to the low SINR, a great amount of transmit power has to be allocated to the weak MU to fulfil the QoS. Hence, it is more sensible to assign MUs with lower QoS requirements as the weak MUs to improve the resource efficiency and to enhance the network throughput. Therefore, we propose a QoS-based clustering scheme, which assigns the MUs with higher or lower QoS requirements as the strong or weak MUs, respectively. 

To be specific, the objective of the QoS-based clustering method is to maximize the minimum QoS deviation among all clusters. The clustering problem can be formulated as $\mathop {\max } \mathop{\min }\limits_{l={1, \cdots, K/2}}(R^{l,s}_{QoS} - R^{l,w}_{QoS})$, 
where $R^{l,s}_{QoS}$ and $R^{l,w}_{QoS}$ denote the QoS requirements of the strong and the weak MUs in the $l$-th cluster, respectively. 

To achieve the maximal QoS deviation, we propose a simple clustering method when each cluster consists of two MUs. We assume that $K$ is an even number and all MUs are ordered in terms of their QoS requirements, namely, the $k$-th MU has the $k$-th highest QoS requirement. The maximal QoS deviation can be achieved by assigning the $k$-th MU and the $(k+K/2)$-th MU into the same cluster, for all $k \leq K/2$.

\subsection{Problem Formulation}

The optimization goal is to maximize the total throughput of the network by jointly optimizing the RIS phase shift ${{\boldsymbol{\theta}}} = [{\theta _{1}}, \cdots ,{\theta _{n}}, \cdots ,{\theta _{N}}]$ and the BS power allocation vector $\mathbf{P}=[\mathbf{P}_s, \mathbf{P}_w]$, where $\mathbf{P}_s = [p_{1,s}, \cdots, p_{K/2,s}]$ and $\mathbf{P}_w = [p_{1,w}, \cdots,p_{K/2,w}]$. The optimization problem is formulated as
\begin{center}
\begin{subequations}\label{eq: short-term opt problem}
\begin{align}
\mathop {\max }\limits_{\boldsymbol{\theta}, \mathbf{P}} {\kern 1pt} {\kern 1pt} {\kern 1pt}  R = \sum\nolimits_{l = 1}^{K/2} (R_{l,s} + R_{l,w}) \label{eq: short-term maxR}\\
{\text{s}}{\text{.t}}{\text{.}}{\kern 1pt} {\kern 1pt} {\kern 1pt} {\kern 1pt} {\kern 1pt} {\kern 1pt} {\kern 1pt} {\kern 1pt} {\kern 1pt} {\kern 1pt} {\kern 1pt} {\kern 1pt} {\kern 1pt} {\kern 1pt} {\kern 1pt} {\kern 1pt} {R_{l,i}} \ge R_{\text{QoS} }^{l,i} ,\forall l, \forall       i \in \{ s,w\} \label{eq: short-term constraint1}\\
\left| {{e^{j\theta_n}}} \right| = 1,\forall n \label{eq: short-term constraint2}\\
\sum\nolimits_{l=1}^{K/2} (p_{l,s} + p_{l,w}) \leq P_{max}\label{eq: short-term constraint3}
\end{align}
\end{subequations}
\end{center}
where $R_{l,i} = B_l\log_2(1+\gamma_{l,i})$ denotes the throughput achieved by MU $i$ in cluster $l$ and $R_{\text{QoS} }^{l,i}$ denotes the minimal QoS requirement of the given MU. Hence, \eqref{eq: short-term constraint1} represents the minimum transmit rate constraint. Moreover, \eqref{eq: short-term constraint2} denotes the phase shift constraint of the RIS and (8d) qualifies the total transmit power constraint of the BS. Due to the non-convex constraint~\eqref{eq: short-term constraint2}, the optimization problem can not be directly solved by conventional approaches. Hence, we proposed to tackle the joint optimization problem utilizing machine learning techniques.

\section{DL-Based Power Allocation and Phase Shift Optimization}\label{sec: DL solution}
In this section, we introduce the proposed meta-learning enabled DL algorithm that jointly optimizes the power allocation and the RIS phase shift, given MUs' QoS requirements.

\subsection{Proposed MAML-based Phase Shift and Power Allocation Optimization Algorithm}

The main idea of DL algorithms is to extensively train a neural network such that, given any inputs, the outputs of the network achieve minimal loss. A conventional design is to construct a neural network that outputs all optimization variables, namely, the RIS phase shift and the power allocation. However, this design will result in an extremely large input space that consists of all channel information and QoS information. In particular, all channel matrices contribute $(2K\times N +2K\times M + 2N\times M)$ to the input dimension, leading to exceedingly expensive computational costs. Moreover, the phase shift and the power allocation have vastly different value ranges and distributions, which greatly increase the training difficulty.

\begin{remark}
Optimizing the phase shift requires the knowledge of all channels among the BS, the RIS and the MUs. However, the optimization of the power allocation only requires the information of the combined channel.
\end{remark}

Inspired by the fact that the combined channel of all MUs, denoted by $\mathbf{H} = [\mathbf{h}_{1,s}, \mathbf{h}_{1,w}, \cdots, \mathbf{h}_{K/2,s}, \mathbf{h}_{K/2,w}]$, provides sufficient channel information for optimizing $\mathbf{P}$ but not $\boldsymbol{\theta}$, we propose to design the neural network to input the combined channel and output the optimized power allocation. The real and the imaginary parts of the combined channel $\mathbf{H}$ contributes $(2K\times M)$ to the input dimension, which is significantly smaller than the input dimension of the intuitive design. The neural network $G_{\eta}$ is formulated as 
\begin{equation}\label{eq: P = G(H, R, theta)}
    \mathbf{P} = G_{\eta}(\mathbf{H}( \boldsymbol{\theta}), \mathbf{R}_{QoS}, \mathbf{L}_{\text{path}}),
\end{equation}
where $\mathbf{H}(\boldsymbol{\theta})$ denotes the combined channel calculated using the phase shift $\boldsymbol{\theta}$, $\mathbf{R}_{QoS}\in \mathbb{R}^{K}$ denotes the QoS requirement vector, and $\mathbf{L}_{\text{path}} \in \mathbb{R}^{K}$ is the path loss vector. The phase shift $\boldsymbol{\theta}$ is optimized separately using a gradient descent algorithm. Two optimization algorithms are connected in an alternating structure, by using the output of the other algorithm as the input. 

In contrast to the conventional alternating optimization approach, we train the neural network $G_{\eta}$ to output the optimized power allocation given any phase shift. Hence, given a trained $G_{\eta}$, we can find the optimized pair of $\boldsymbol{\theta}$ and $\mathbf{P}$ by solely performing the optimization on $\boldsymbol{\theta}$. To further improve the convergence rate of the gradient descent algorithm, the network $G_{\eta}$ is trained using MAML, such that, the optimized pair of $\boldsymbol{\theta}$ and $\mathbf{P}$ can be obtained in as few as five iterations.


MAML is a meta-learning technique, which is designed to optimize the
model parameters such that a few gradient steps will produce
a maximally effective performance on a new task~\cite{few_shot1_MAML}.
As demonstrated in~\cite{DCS_DeepMind}, MAML can be employed to reduce the number of gradient descent steps required to optimize the network input space. 
In our model, the network inputs are optimized by adjusting the phase shift $\boldsymbol{\theta}$. Moreover, the gradient descent steps on $\boldsymbol{\theta}$ are performed by back-propagating through the weights of $G_{\eta}$. Hence, we propose to train $G_{\eta}$ with MAML to obtain a set of network weights that can greatly reduce the number of gradient steps required to update $\boldsymbol{\theta}$.

\subsection{Loss Functions}
As described in~\eqref{eq: short-term opt problem}, both $\boldsymbol{\theta}$ and $\mathbf{P}$ need to be optimized to maximize the system throughput given the constraints. Hence, they share the same loss function, denoted by $\mathcal{L}(\boldsymbol{\theta},\boldsymbol{\eta})$, where $\boldsymbol{\eta}$ is the weights of the neural network $G_{\eta}$ that outputs $\mathbf{P}$. The loss function consists of two parts, the total throughput and the constraint term for enforcing the QoS requirements, given by
\small{\begin{multline}\label{eq: loss function L()}
    \mathcal{L}(\boldsymbol{\theta}, \boldsymbol{\eta}) = w_{1} \sum_{l=1}^{K/2}\sum_{i=s,w} R_{l,i}(\boldsymbol{\theta}, \boldsymbol{\eta}) +\\ w_{2} \sum_{l=1}^{K/2}\sum_{i=s,w} \text{max}\big(R_{l,i}(\boldsymbol{\theta}, \boldsymbol{\eta}) - R^{l,i}_{QoS}, 0\big),
\end{multline}}
\normalsize
where $R_{l,i}(\boldsymbol{\theta}, \boldsymbol{\eta})$ is the sum rate calculated using $\boldsymbol{\theta}$ and $ \boldsymbol{\eta}$, and $\text{max}\big(R_{l,i}(\boldsymbol{\theta}, \boldsymbol{\eta}) - R^{l,i}_{QoS}, 0\big)$ indicates the QoS deficiency of MU $i$ in cluster $l$. The weights $w_1$ and $w_2$ are tuned during training. Since the sum rate is positive-valued and the QoS deficiency is negative-valued, $w_1$ and $w_2$ should be negative and positive, respectively. 

Suppose we aim to optimize $\boldsymbol{\theta}$ in $J$ gradient steps, we can derive the gradient descent formula in the $j$-th gradient step of the $p$-th training episode as
\begin{equation}\label{eq: phase shift update equation}
    \boldsymbol{\theta}^{(j)} \leftarrow \boldsymbol{\theta}^{(j-1)}- \gamma_\theta \frac{\partial}{\partial \boldsymbol{\theta}^{(j-1)}} 
   \mathcal{L}(\boldsymbol{\theta}^{(j-1)},\boldsymbol{\eta}^{(p-1)}),
\end{equation}
where $\gamma_\theta$ denotes the step size and $\boldsymbol{\eta}^{(p-1)}$ denotes the neural network weights obtained in the previous training episode. In order to satisfy the phase shift constraint in \eqref{eq: short-term constraint2}, we clip the values of $\boldsymbol{\theta}$ to $[0, 2\pi]$ after each update. Based on \eqref{eq: phase shift update equation}, the loss function after completing the $J$-th gradient step is therefore $\mathcal{L}(\boldsymbol{\theta}^{(J)}, \boldsymbol{\eta}^{(p-1)})$. Then, we optimize the neural network $G_{\eta}$ to further minimize $\mathcal{L}(\boldsymbol{\theta}^{(J)}, \boldsymbol{\eta}^{(p-1)})$ through the following update formula
\begin{equation}\label{eq: eta update equation}
    \boldsymbol{\eta}^{(p)} \leftarrow \boldsymbol{\eta}^{(p-1)}- \gamma_\eta \frac{\partial}{\partial \boldsymbol{\eta}^{(p-1)}} 
   \mathcal{L}(\boldsymbol{\theta}^{(J)}, \boldsymbol{\eta}^{(p-1)}),
\end{equation}
where $\gamma_\eta$ denotes the learning rate.


\begin{algorithm}[t]
\caption{Meta-learning Based Training Algorithm}
\begin{algorithmic}[1]
\label{Algorithm: train}
    \renewcommand{\algorithmicrequire}{\textbf{Input:}}
    \renewcommand{\algorithmicensure}{\textbf{Output:}}
    \REQUIRE Channel matrix $\mathbf{H}$, QoS vector $\mathbf{R}_{QoS}$, MU locations, neural network $G_{\eta}$, number of phase shift update steps $J$, phase shift learning rate $\gamma$
    \ENSURE  Trained neural network $G_{\hat{\eta}}$
 \\ Initialize $\boldsymbol{\eta}$
    \REPEAT
        \FOR {each episode} 
            \STATE Initialize phase shift $\theta_1, ..., \theta_N \overset{\text{iid}}{\sim} \mathcal{U}(0, 2\pi)$
            \STATE Calculate path loss vector $\mathbf{L}_{\text{path}}$
            \FOR {$j = 0$ to $J-1$}
                \STATE Obtain power allocation $\mathbf{P}^{(j)} = G_{\eta}(\mathbf{H}(\boldsymbol{\theta}^{(j)}), \mathbf{R}_{QoS}, \mathbf{L}_{\text{path}})$
                \STATE Calculate loss function $\mathcal{L}(\boldsymbol{\theta}^{(j)} ; \boldsymbol{\eta})$
                \STATE Update phase shift using \eqref{eq: phase shift update equation}
            \ENDFOR
        
        \STATE Given the optimized phase shift $\boldsymbol{\theta}^{(J)}$, calculate the optimized power allocation $\mathbf{P}^{(J)} = G_{\eta}(\mathbf{H}(\boldsymbol{\theta}^{(J)}), \mathbf{R}_{QoS}, \mathbf{L}_{\text{path}})$
        \STATE Calculate loss function $\mathcal{L}(\boldsymbol{\theta}^{(J)}, \boldsymbol{\eta})$ using the optimized phase shift
        \STATE Update network weights using \eqref{eq: eta update equation}
        \ENDFOR
    \UNTIL{reaches the maximum training steps}
    \STATE \textbf{Return} $G_{\hat{\eta}}$
\end{algorithmic} 
\end{algorithm}

To employ MAML, we calculate~\eqref{eq: eta update equation} by implicitly performing the second order differentiation with respect to the loss function $\mathcal{L}(\boldsymbol{\theta}^{(J)}, \boldsymbol{\eta}^{(p-1)})$ and back-propagating through the $J$ phase shift optimization steps in \eqref{eq: phase shift update equation}. What's more, we also perform MAML on the hyper-parameter $\gamma_\theta$ against the loss function $\mathcal{L}(\boldsymbol{\theta}^{(J)}, \boldsymbol{\eta}^{(p-1)})$ to reduce the need for further hyper-parameter tuning. The update equation of $\gamma_\theta$ can be derived in the same way as in~\eqref{eq: eta update equation}. 

\begin{figure}[t]
    \centering
    \includegraphics[width=0.4\textwidth]{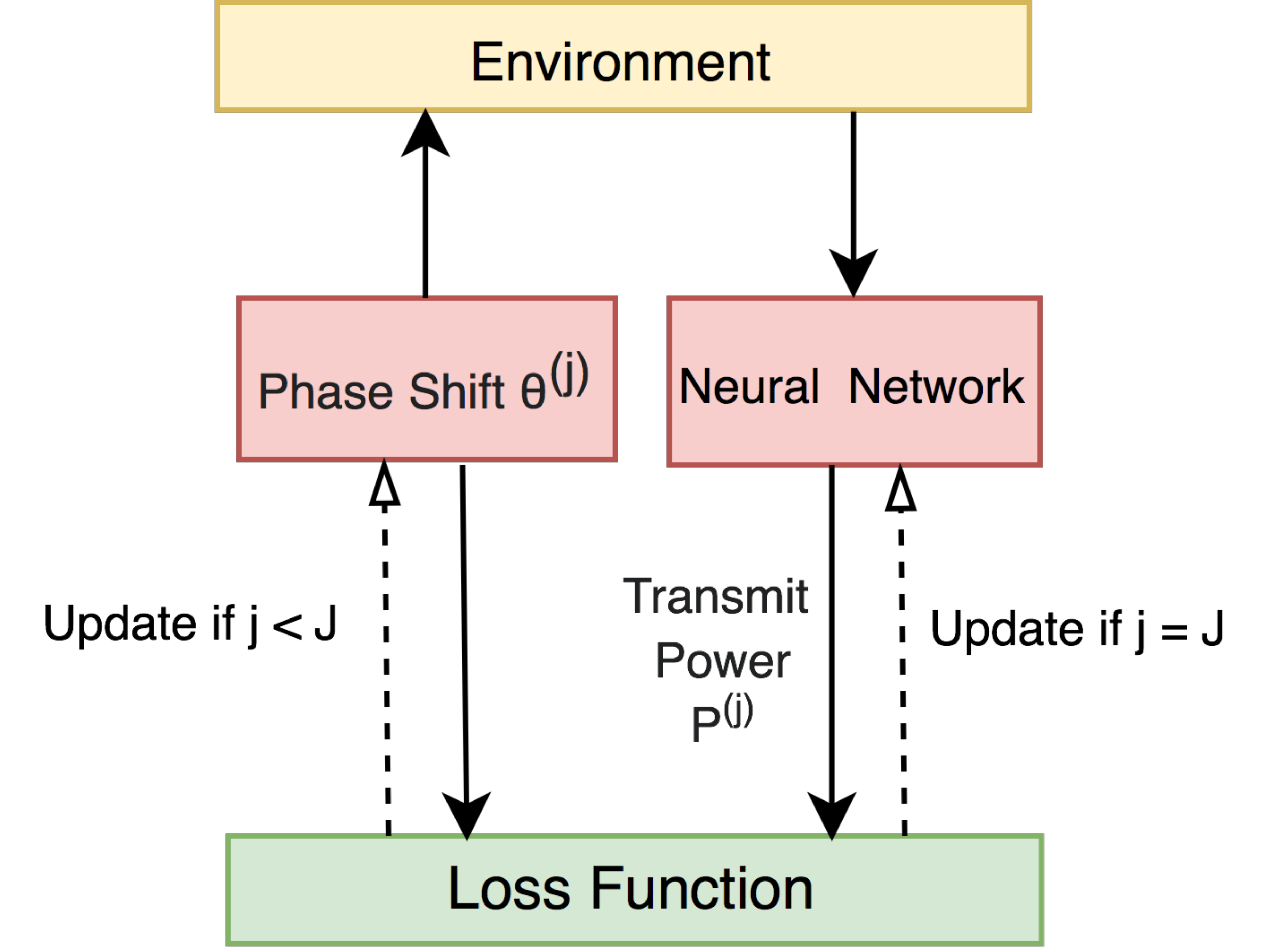}
    \caption{An illustration of the MAML-based training framework.}
    \label{fig: DL_network_figure}
\end{figure} 

\subsection{Training Algorithm}
As shown in Fig.~\ref{fig: DL_network_figure}, each training epoch can be divided into two stages, corresponding to the inner and outer MAML steps:
\begin{enumerate}
    \item \textit{Phase shift optimization (inner step)}: The initial phase shift is sampled according to a random uniform distribution, i.e. $\boldsymbol{\theta}^{(0)} \sim \mathcal{U}(0, 2\pi)$. In the $j$-th gradient loop, the corresponding power allocation $\mathbf{P}^{(j)}$ is obtained based on~\eqref{eq: P = G(H, R, theta)}, using $\boldsymbol{\theta}^{(j)}$. Then, $\boldsymbol{\theta}^{(j)}$ is optimized with respect to the loss function $\mathcal{L}(\boldsymbol{\theta}^{(j)}, \boldsymbol{\eta})$, as in~\eqref{eq: phase shift update equation}. We repeat~\eqref{eq: phase shift update equation} for $J$ iterations. The final optimized phase shift is thus $\boldsymbol{\theta}^{(J)}$. 
    \item \textit{Power allocation optimization (outer step)}: After completing $J$ gradient descent loops, the current optimal power allocation $\mathbf{P}^{(J)}$ can be computed using $\boldsymbol{\theta}^{(J)}$ and~\eqref{eq: P = G(H, R, theta)}. Then, the network weights $\boldsymbol{\eta}$ are updated according to~\eqref{eq: eta update equation}, by backpropagating through all $J$ gradient descent iterations.
\end{enumerate}


The pseudocode of the training algorithm is presented in Algorithm~\ref{Algorithm: train}, where lines 2-8 correspond to the phase shift optimization steps and lines 9-11 correspond to the power allocation optimization steps. To apply the trained network on test datasets, we only need to perform the phase shift optimization procedure for $J$ times, after which the optimized phase shift $\boldsymbol{\theta}^{(J)}$ and the corresponding power allocation $\mathbf{P}^{(J)}$, are the solutions to our joint optimization problem in~\eqref{eq: short-term opt problem}.

\subsection{Complexity Analysis}
The computational complexity of the proposed joint optimization algorithm mainly depends on three factors, namely, the number of phase shift update steps $J$, the complexity of the loss function, and the complexity of the neural network. Trivially, the complexity of~\eqref{eq: loss function L()} is dominated by the calculation of individual MU's combined channel vector $\mathbf{h}^H_{l,i}$, of which the complexity is $\mathcal{O}(NM)$. Since we compute the combined channel for each MU, the total complexity induced by calculating the loss function is $\mathcal{O}(NKM)$. Then, we note that a fully-connected neural network of $D$ layers, including input and output layers, has a computational complexity of $\mathcal{O}(\sum_{i=1}^{D-1} n_i n_{i+1})$, where $n_i$ is the number of neurons in layer $i$. Therefore, the proposed algorithm has a complexity of $\mathcal{O}(JNKM\sum_{i=1}^{D-1} n_i n_{i+1})$.


\section{Simulations}\label{sec:experiment}
In this section, we present the simulation results for the proposed MAML-based optimization algorithm for RIS-assisted NOMA downlink networks. We consider MUs moving in a square area of width 10 meters. The BS is located at a corner of the area and the RIS is randomly placed in the area. In all simulations, the path loss exponent is set to $\alpha=3$, the noise power spectral frequency is -169 dBm/Hz, and the total bandwidth is 4 MHz. The number of phase shift optimization steps is $J=5$ unless otherwise stated.

\begin{figure}[t]
\centering
    \includegraphics[width=0.4\textwidth]{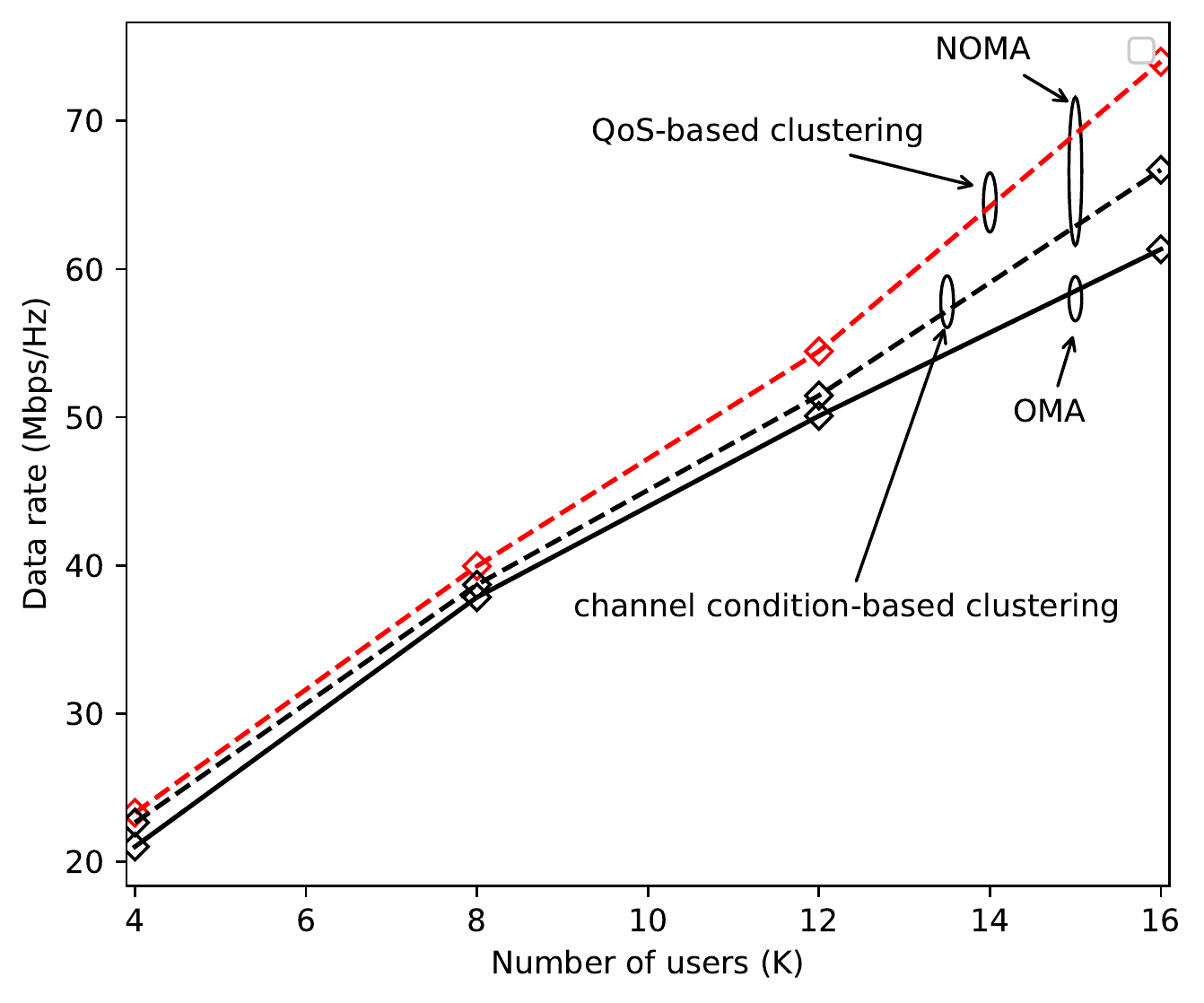}
    \caption{Sum rate versus the number of MUs, $K$, in NOMA and OMA systems using the QoS-based or the channel condition-based clustering schemes.}
    \label{fig: DL_DRL_rate_vs_K_ST}
\end{figure}

\begin{figure}[t]
\centering
    \includegraphics[width=0.4\textwidth]{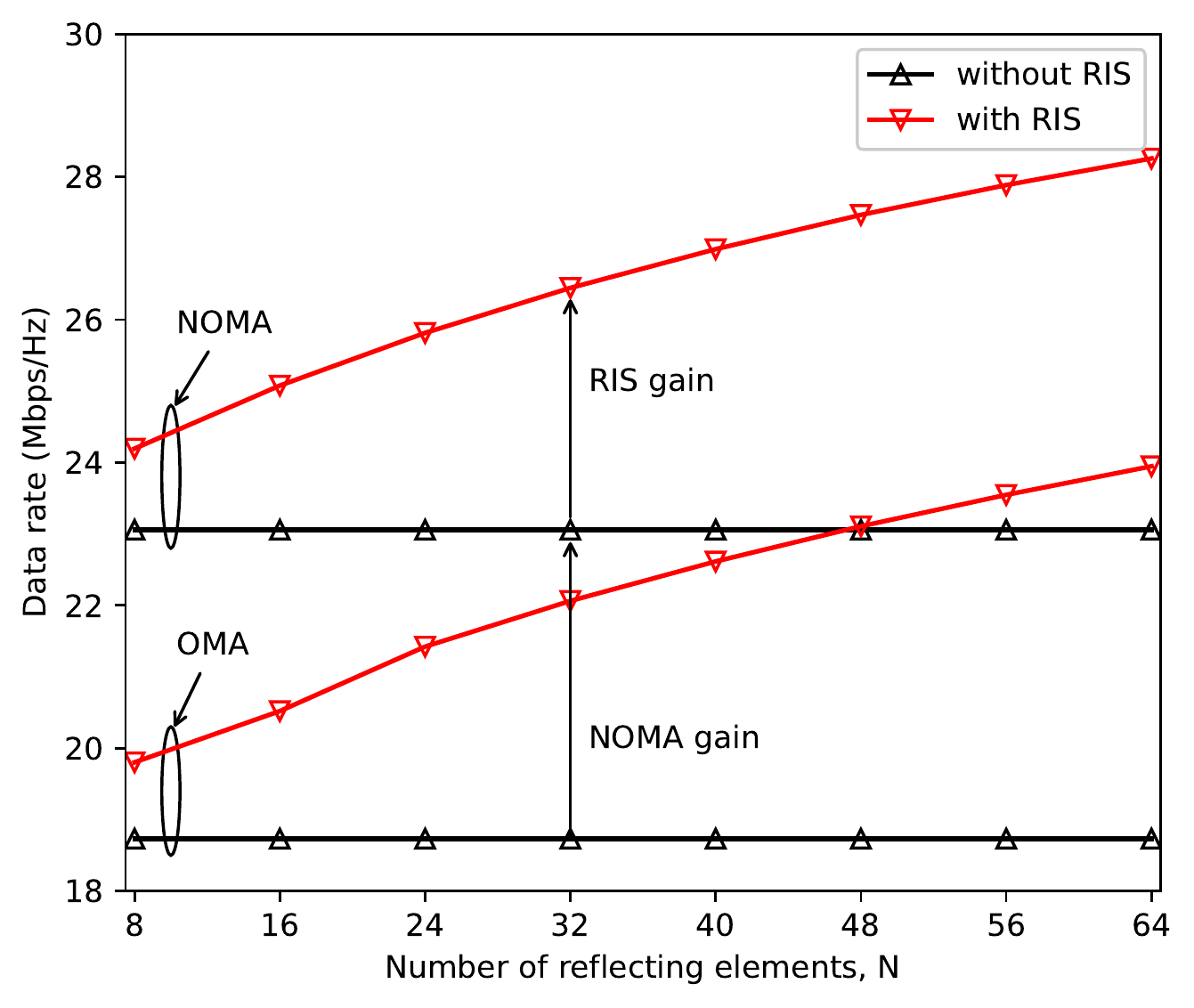}
    \caption{Sum rate versus the number of reflecting elements $N$ for NOMA and OMA cases, given $M=16$ antennas and 20 dBm transmit power at BS.}
    \label{fig: rate_vs_N}
\end{figure}

\begin{figure}[t]
\centering
    \includegraphics[width=0.4\textwidth]{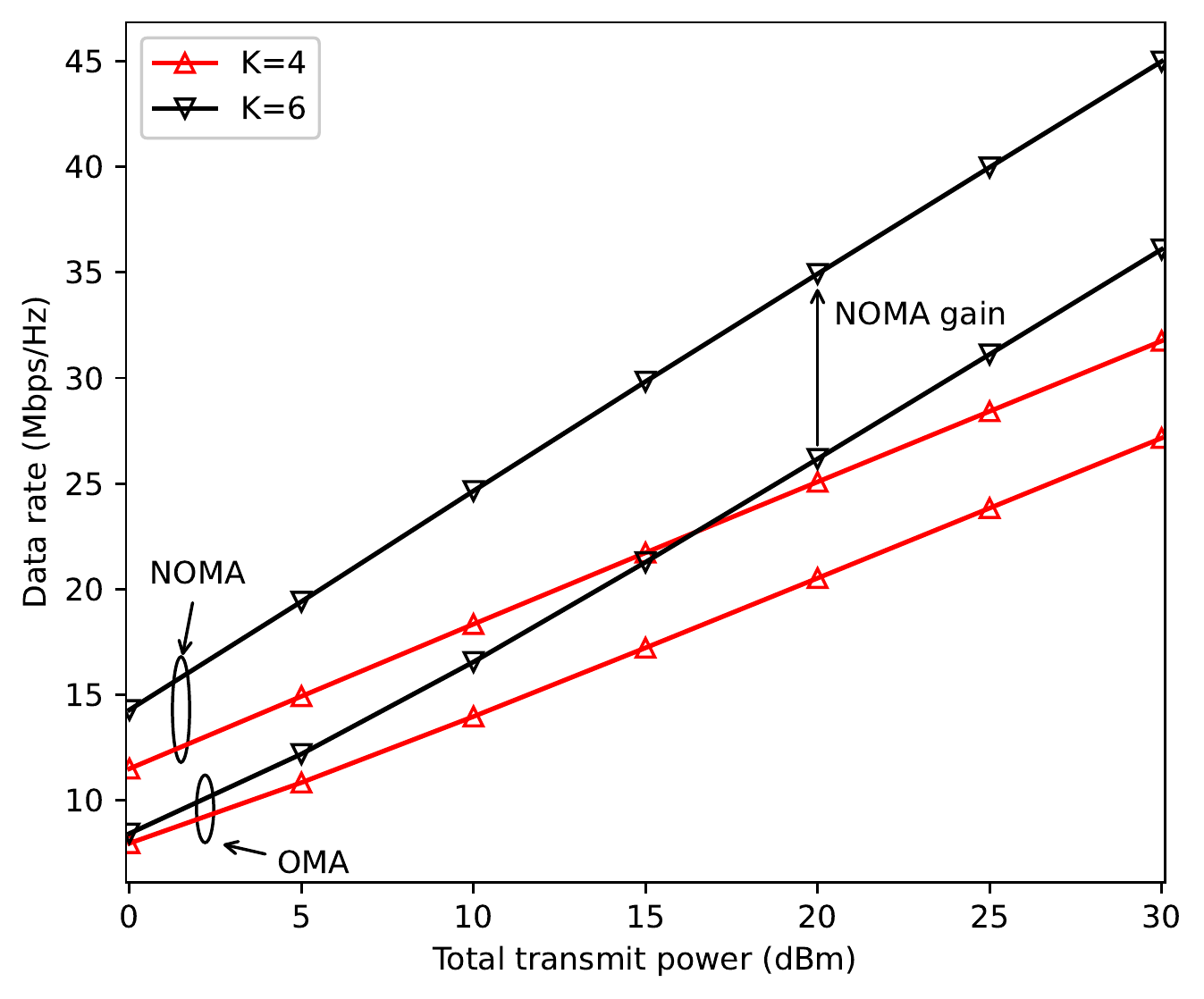}
    \caption{Sum rate versus total transmit power at BS for NOMA and OMA cases, given $M=16$ BS antennas and $N=16$ reflecting elements.}
    \label{fig: rate_vs_power}
\end{figure}


\subsection{QoS-based versus Channel Condition-based Clustering Methods}
In Fig.~\ref{fig: DL_DRL_rate_vs_K_ST}, we compare the network throughput given different clustering methods, namely, the proposed QoS-based clustering scheme and the conventional channel condition-based clustering scheme. Simulations are performed with $M=24$ BS antennas and $P_{\text{max}} = 20$ dBm maximum transmit power. It can be observed that, for smaller numbers of MUs, i.e., $K\leq 8$, the performance difference between two clustering schemes is small because their clustering results are likely to be similar. However, when there are more MUs in the system, i.e., $K\geq 12$, the proposed QoS-based method starts to achieve higher throughput than the conventional channel condition-based approach. The performance gains further increases as the number of MUs increases. 

\subsection{Sum Rate versus the Number of RIS Elements}
In Fig.~\ref{fig: rate_vs_N}, we observe that the proposed NOMA network with the QoS-based clustering scheme outperforms the conventional OMA network of around 4 dBm/Hz of sum rate, without the enhancement of RIS. In both NOMA and OMA networks, the deployment of RIS induces approximately 5\% to 25\% throughput gain as the number of reflecting elements ranges from $N=8$ to $N=64$. Higher performance gain can be attained by increasing the number of RIS elements, however, the cost of optimization complexity and the deployment cost increases as well.

\subsection{Sum Rate versus BS Total Transmission Power}
Fig.~\ref{fig: rate_vs_power} illustrates the throughput performance between OMA and NOMA systems as the BS power varies between 0 dBm and 30 dBm. It can be observed that the NOMA system outperforms the OMA system for all values of BS transmit power, given the same number of MUs. We also notice that the NOMA network with 4 MUs achieves higher throughput than the OMA network with 6 MUs when the BS power is less than 15 dBm. Moreover, as the number of MUs increases, the throughput of the NOMA networks increases by a larger amount compared to the throughput of the OMA networks.


\section{Conclusions}\label{sec:conclusion}
In this article, we proposed a QoS-based clustering method to improve the resource efficiency in RIS-assisted NOMA networks. We formulated the sum rate maximization problem by jointly optimizing the RIS phase shift and the BS power allocation. The proposed DL solution utilized a low-complexity network architecture and employed MAML to improve the convergence rate. Simulation results demonstrated that higher transmission throughput was achieved by the proposed QoS-based clustering method than the baseline method. Results also illustrated that the proposed QoS-based NOMA model achieved higher throughput compared to the conventional OMA models for a wide range of values of the BS transmit power and the RIS elements. Moreover, the implementation of RIS improved network throughput by a substantial amount, which further grows as the number of reflecting elements increases. 

\bibliographystyle{IEEEtran}
\bibliography{Yixuan_GC_2021}

\end{document}